\documentstyle[11pt,aaspp,tighten]{article}
\newcommand{\as}{$^{\prime\prime}~$}
\newcommand{\am}{$^{\prime}~$}

\begin{document}

\title{\bf  Two X-ray Sources in 30 Doradus:
Wolf-Rayet + Black Hole Binaries?}

\author{Q. Daniel Wang}
\affil{Dept. of Physics \& Astronomy, Northwestern University}
\affil{ 2145 Sheridan Road, Evanston,~IL 60208-3112}
\affil{Electronic mail: wqd@nwu.edu}

\begin{abstract}

	I report the detection of two X-ray sources of luminosities
$\sim 10^{36} {\rm~ergs~s^{-1}}$ in the central region of 30 Doradus. These
two sources appear point-like in images taken with the {\sl ROSAT} HRI.
One of the sources is most likely associated with a close spectroscopic binary
R140a2 (WN6) with an orbital period of 2.76 days. The mass of the unknown
binary component is likely in the range of $2.4 - 15{\rm~M}_\odot$. This
suggests that  the X-ray source could represent a long-sought class
of binaries containing a Wolf-Rayet star and a black hole. The other source,
which coincides spatially with  Mk34 (WN4.5), may have a similar nature.

	Available X-ray spectral data support this Wolf-Rayet + black-hole
binary explanation of the two sources. I have used a {\sl ROSAT} PSPC
observation to show that the sources have ultrasoft spectra with possible
intrinsic absorption. Modeled with multicolor blackbody disks, the spectra
provide estimates of the disks' characteristic inner radii,
which are in agreement with those obtained for the known
black-hole candidates.  An {\sl ASCA} observation has further
revealed a hard X-ray spectral component from the central 30 Doradus region.
This component, represented by a power law with a photon index
of $\sim 2.4$, may belong to the two sources. The characteristics
of both the power-law and ultrasoft components strongly indicates
that the two sources are black-hole candidates in high-mass
X-ray binaries.

\end{abstract}
%\clearpage
\section{Introduction}

	Although 30 Doradus in the Large Magellanic Cloud (LMC)
has been a region of intense interest and study for a long time
(Malumuth \& Heap 1994 and references therein), detailed
X-ray investigations of the region started only recently. Using data from the
{\sl Einstein} Imaging Proportional Counter, Wang \& Helfand (1991; hereafter
WH91) have shown that diffuse hot gas is the dominant source of X-rays
from the region. The hot gas occupies an area of dimension $\sim 300$~pc
around R136, the core of the central cluster NGC~2070
(15~pc = 1\am at an adopted distance of 51~kpc). But
little has yet been found about the discrete X-ray source population in the
region. The {\sl Einstein} High Resolution Imager (EHRI) has indicated
(only at $\sim 3\sigma$ levels) the presence of two X-ray sources
 in the central region near R136 (WH91).

	To study both the discrete sources and diffuse hot gas in the region,
I obtained a deep X-ray image with the {\sl ROSAT} High Resolution
Imager (hereafter RHRI; Pfeffermann et al. 1988). This image has led to
a firm detection of these two sources and an identification of their probable
optical counterparts. I have further extracted information on the sources
from  the {\sl ROSAT} archive, an {\sl ASCA} spectrum, and various optical
observations.  I conclude that the two sources are most likely
high mass X-ray binaries containing black-hole candidates (HMXB BHCs).
In the rest of this paper, I first describe properties of the observations
and data analysis,  dealing primarily with
astrometric problems (\S~2). Then in \S~3, results are presented
on spatial, spectral, and timing properties of the sources.
The nature of the sources is explored by combining information
from various observations and by comparing with other X-ray-emitting
objects (\S~4). I give a summary in \S~5.
All statistical uncertainties are quoted at the 90\% confidence level.

\section {Observations and Data Analysis}

	While an up-to-date description of {\sl ROSAT}
can be found in Briel et al. (1994), I quote here only the most
relevant information about the telescope/RHRI system and concentrate on
the primary observation ({\sl ROSAT} sequence No. rh600228) used in this work.
The  observation was targeted at R136 (RA $= 5^h 38^m 43\fs2;$
Decl.$ = -69^\circ 6^\prime $0\as --- J2000) and accumulated  a total live
time of 30153~s over various time segments during 1992 December-10 to 1993
June-14. The field of view of the RHRI system was $\sim 35^\prime$ diameter.
The spatial resolution, if characterized by a point-spread function (PSF), was
about 6\as (FWHM) on axis and degraded to $\sim 30^{\prime\prime}$
at $15^\prime$ off axis. The system had a mean effective area of
$\sim 80{\rm~cm^2}$, a factor of $\sim 10$ greater than that of the EHRI,
 over the 0.65 - 1.9~keV range (where the system's effective area was larger
than 20\% of its peak value $\sim 95 {\rm~cm^{-2}}$ at $\sim 1.1$~keV).
At $\lesssim 0.5$~keV (i.e., around a second peak of the system's effective
area at $\sim 0.28$~keV), the 30 Doradus contribution can be neglected
because of interstellar absorption along the line of sight.

	Discrete X-ray sources in the field were detected and analyzed using
a maximum likelihood algorithm. This algorithm uses a detection aperture that
varies according to the  size of the PSF as a function of off-axis angle.
Otherwise, the algorithm is very similar to that used in the Standard Analysis
Software System (SASS; Downes et al. 1992; Briel et al. 1994).

	The aspect error of a {\sl ROSAT} observation is typically a few
arcseconds, but occasionally exceeds more than 10\as. I reduced this error
in my observation  by comparing X-ray and optical positions of three
identified point-like X-ray sources in the field. The optical positions
(Table 1) were adopted from the Guide Star Catalog for cal~69 (V = 12) and
cal~71 (V = 10), and from Shaerer et al. (1994) for the 50~ms pulsar.  This
position comparison led to a plate solution, which required shifting the
observation 1\as to the west and 3\as to the south from the intended pointing
direction, plus a rotation of 0\fdg42 clockwise
around the pointing axis, and gave a mean SASS pixel size of 0\farcs498.
The required rotation  agrees with the systematic value
$0\fdg4\pm0\fdg1$ discovered by Kuerster (1993).
The mean pixel size is  consistent with the value
$0\farcs499 \pm 0\farcs001$, obtained from a plate solution of
an RHRI observation on M31 (Briel et al. 1994 and references therein).

	The uncertainty in such a plate solution is primarily systematic.
The most important fact that I did not account for
is probably the azimuthal asymmetry of the PSF at large off-axis
angles. The asymmetry causes a systematic offset of a source centroid
 (which is the source's detection position)
from the source's true position towards the pointing axis. This offset
 introduces a position error up to $\sim 10$\as near the edges of an
observation, although its exact dependance on source off-axis angle has yet
to be quantified. The reference sources I used are all at large off-axis
angles (Fig.~1 and Table 1). Thus, accounting for the effect would make the
mean SASS pixel size smaller ($\sim 0\farcs001$). Fortunately,
the reference sources  are located on opposite sides relative to the axis;
much of the effect on estimating the shifts and the rotation was canceled
in the solution. But one may still expect an error up to $\sim$ 3\as in the
astrometry of the observation, judging from the differences between X-ray and
optical positions (Table 1) and between solutions obtained from selecting
different pairs of the reference stars.

	Moreover, proper motions of the two Galactic stars may produce
position shifts of $\sim 1^{\prime\prime}$ over a period of $\sim 13$~yrs
between the ESO survey (from which two reference stars' optical positions
were derived) and the RHRI observation. For this reason, I compared
Guide Star Catalog  positions of stars in an H$\alpha$ image of 30 Doradus
(Smith et al. 1992) taken within  two years of the X-ray observation,
and found no noticeable  position
deviation ($\lesssim 1^{\prime\prime}$) of cal~71 relative to other
stars in the field. Cal~69 is not in the field of the image, but
excluding the star has little effect on the results of the plate solution.

	I further corrected for time-dependent aspect errors in the
observation. First, for  individual observing segments that were separated by
more than a hour, I calculated the position centroids of the three brightest
sources in the field (i.e., cal~71, 50~ms pulsar, and N157B).  Located
near the southern edge of the field, cal~69 is not used to
avoid effects caused by the spacecraft wobble. I estimated  the pointing
errors, using a $\chi^2$ fit to the  deviations of
the source centroids in each individual segment from
the sources' mean positions over the entire observation.
Only large absolute errors ($\geq 2\farcs5$) were corrected;
those smaller ones were less statistically significant.

	Similar astrometric corrections were also made on the RHRI observation
(wh500036), which was originally taken by B. E. Aschenbach and was
downloaded from the {\sl ROSAT} archive.
This observation, pointed at the nearby Crab-like pulsar N157B (Table 1),
had a live time of 10910~s. The observation was shifted:
$0\farcs5$ to the west, 3\farcs0 to the south, and 0\fdg42
clockwise around the pointing axis, similar to the corrections
made on rh600228.

	All these astrometric corrections were made directly
on the X-ray source positions and on the sky positions of individual
RHRI counts before being cast into images (e.g., Figs.~1-3).

	To examine physical properties of the two X-ray sources, I extracted
spectral data from an observation taken by Y-H. Chu (1993) with the
{\sl ROSAT} Positional Sensitive Proportional Counter (PSPC)
during 1992 April-29 to May-18. The PSPC had about
six independent energy bands over the 0.1 to 2 keV range.
For each source, I obtained a spectrum from a pulse height-invariant
(PI) channel distribution of net source counts.
(The PI binning uses the pulse height energies
of individual events recorded by the PSPC and incorporates both the
instrument temporal and spatial gain corrections.) The net source counts
were obtained from on-source counts from a circle of
0\farcm4 radius around the source centroid, minus the area-normalized
background counts from a concentric annulus
with an inner radius of 1\farcm2 and an outer radius of 2\farcm5.
Vignetting corrections were applied to each of the counts. The
contamination from the other source in the background annulus
was removed by excluding the area
within the source's 90\% radius (defined to contain
 90\% of the counts of a point-like source).

\section { Results}

	 Fig.~1 is an overlay of X-ray contours on a digitized ESO plate,
illustrating the overall correspondence between X-ray and optical
objects in the 30 Doradus region. Fig.~2 is a representation of the X-ray
image, optimally smoothed to reveal faint diffuse X-ray features. These two
figures confirm the {\sl Einstein} results on the global morphology of
diffuse X-rays emanating from the 30 Doradus nebula (WH91); the X-ray-emitting
gas is clearly enclosed by filaments, or loops and shells, of warm ionized gas.
The wealthy details of the diffuse X-ray component and
 N157B will be discussed  elsewhere. Here, I concentrate
on discrete sources in the central region.

\subsection {Spatial Properties}

  	Table 1 is a list of point-like sources detected  in the 30 Doradus
field (Fig.~1). Sources 5 and 7 are the only two sources that
show dominant point-like components within the entire nebula above a $5\sigma$
 source detection threshold of $\sim 3 \times 10^{-3} {\rm~counts~s^{-1}}$.
I compared the observed radial intensity profiles of the two
sources with the PSF of the RHRI system. $\chi^2$ fits were performed
with the normalization of the PSF and the intensity of an assumed uniform
local background as two free parameters. Source 5 is consistent with being
point-like ($\chi^2/n.d.f. = 15.8/13$), but Source 7 clearly contains a second
contribution in addition to a point-like component (28.0/13).

	In Figs.~3-4, one can see an intensity excess between $\sim$
5\as - 15\as off the point-like centroid
of Source 7. This excess extends primarily in the direction of R136,
covering an area of the greatest massive star concentration in the region.
The excess accounts for $\sim 10\%$ of the count rate of the source,
corresponding to an {\sl unabsorbed} luminosity of
$4 \times 10^{34} {\rm~ergs~s^{-1}}$ in the 0.5 - 2~keV band (see \S~3.2).
Within a 5\as radius of R136, stars with a total $L_{bol} \approx 4
\times 10^7 L_\odot$ (Malumuth \& Heap 1994), are nominally
expected to contribute $0.9 - 5 \times 10^{34}{\rm~ergs~s^{-1}}$, where
the lower limit is derived by assuming $L_x(0.5 - 2~{\rm keV})/L_{bol} \approx
6 \times 10^{-8}$ for Wolf-Rayet (WR) stars (Pollock 1987) and the
upper limit assuming $3 \times 10^{-7}$ for O/B stars (Chlebowski 1989).
Thus, the excess is consistent with the
nominal stellar X-ray contribution in the region, and there is no
evidence that R136 contains an exotic X-ray-emitting object.

	The centroids of Sources 5 and 7 coincide spatially with R140 and
Mk34 (Fig.~3 and Table 1). The optical positions, obtained in a plate solution
of the CCD image presented in Fig.~3, are accurate to $\lesssim
1^{\prime\prime}$ (Parker 1993). Within $\sim 3$\as radii around the centroids
(the adopted systematic X-ray position uncertainty), no obvious alternative
objects are present as optical counterparts of the X-ray sources.
If the position coincidence represents their true physical association
with the optical objects, the sources are then
the first X-ray detection of WR stars (or systems) outside
the Milky Way.

\subsection {Spectra and Luminosities}

	Table 2 summarizes results of model fits to the PSPC spectral data
(Fig.~5). All the three spectral models in the table provide satisfactory
fits for Source 5, but can be rejected for Source 7 at about the 99\%
confidence level. As illustrated in Fig.~5b, a couple of line features
might be present in the spectrum of Source 7
and are not accounted for by the  models. Furthermore, compared to
Sources 5, there might be a hard tail in the spectrum of Source 7.
This may be explained by the fact that the spectrum
includes two components spatially resolved by the RHRI
observation (\S~3.1); the tail could be due to relatively hard X rays
from normal stars. Nevertheless, the overall spectral shapes of Sources 5 and 7
appear similar, being very  steep in the 1-2~keV range and
showing cutoffs near 1~keV. The steep spectra are an indication
of optically-thick emission from accretion disks (see also \S~4).
Therefore, among the models included in Table 2, the multicolor blackbody
disk (MBD) is probably the most realistic representation of the X-ray
spectral data. From the normalizations of the fits with MBD, the inner radii
of the disks are derived (Table 3).

	The characterization of the X-ray spectra provides the conversion
between count rates and energy fluxes of the sources. The {\sl absorbed}
energy fluxes of Sources 5 and 7 are 1.2 and 2.3$\times 10^{-13}
{\rm~ergs~s^{-1}~cm^{-2}}$ in the 0.5 - 2~keV band,
which are insensitive to the choice of a specific spectral model.
Assuming the best-fit MBD parameters,
the luminosities of the two sources are $\sim 7 $ and 6 $ \times 10^{35}
{\rm~ergs~s^{-1}}$ in the 0.5 - 2~keV band.

\subsection {Variability}

	The two X-ray sources are detected in all the three {\sl ROSAT}
observations with no evidence of dramatic aperiodic intensity variation,
and are therefore not X-ray transients. Only Source 5 showed intensity
variation of marginally significance between the two RHRI detections;
the mean source count rate during wh500036 was $\sim 2.0 \pm 0.9$
lower than that during rh600228. No significant variability
was detected during individual observations. But, owing to the low count
rates of the sources, this does not exclude luminosity variations of
$\sim$ 40\%, for an example, on a timescale of 4000~s for both sources.

There is no evidence for a phase-dependent variation in the intensity of
Source 5. A constant intensity fit to the source's RHRI light curve, folded
according to the orbital period of R140a2 (see \S~4.4),
gives $\chi^2/n.d.f = 10.2/8$. The upper limit to the pulse fraction of
the source intensity is $\sim 50\%$, assuming a sinusoidal waveform
as may be expected from a stellar-wind-driven X-ray source in
an eccentric orbit.

\section  {Nature of the X-ray Sources}

	Using the {\sl ROSAT} observations, I have shown that
two luminous X-ray sources are present in the central region of 30 Doradus.
These two sources are most likely associated with WR stars, have X-ray
luminosities of $\sim 10^{36} {\rm~ergs~s^{-1}}$ in the 0.5-2~keV band,
are not X-ray transients, and show steep spectra over the 1-2~keV range and
possible intrinsic X-ray absorption. Combining  results from an observation
made with the {\sl ASCA} X-ray observatory, and information from other
wavelength bands, I now attempt to determine the nature of the sources.

\subsection {Association with Wolf-Rayet Stars}

	The association of the two X-ray sources with the WR stars
appears to be real. 30 Doradus houses $\sim 20$ WR stars
and $\sim 60$ O stars of type hotter than O5 in the central diameter of
$\sim 3^\prime$ around R136 (e.g., Walborn 1991; Parker 1993). Many of these
stars are clustered  into groups such as R140 (see \S~4.4). Statistically,
the chance is small for random position coincidences ($\lesssim
2^{\prime\prime}$) of the WR stars with the only two X-ray sources
in 30 Doradus, and is even smaller for random alignments of
the stars with background or foreground objects
of comparable or greater X-ray intensities.

	The association, if real, provides insight into the nature
of the X-ray sources. Since X-rays from individual stars cannot account for
the observed source intensities (\S~3.1), the most appealing model involves
accretion-powered compact objects, such as black holes (BHs) or neutron stars.
It is believed that most stars near R136 were born during a major star
formation burst about $3\pm1$~Myr ago (e.g., Hyland et al. 1992; Lattanzi et
al. 1994). Therefore, only the most massive stars ($\gtrsim 50 {\rm~M}_\odot$)
 formed during this episode  may have collapsed; the end products are probably
BHs if stellar mass BHs do exist and are formed from stars.
When such a collapse happens in a massive close binary,
the BH is retained in the binary orbit in most cases. Thus,
the two X-ray sources may represent HMXBs,
most probably WR+BH binaries, which are expected as an evolutionary stage of
massive binaries (e.g., Moffat \& Seggewiss 1979).

	The large X-ray absorption indicated in the spectral fits
 (Table 2) could be contributed by stellar winds of the WR stars,
in addition to the interstellar medium along the lines of sight. Interstellar
gas column densities may be estimated from optical extinctions towards
R140b and Mk35, whose optical colors,
not blended by nearby stars, were measured by Parker (1993).
R140b (\S~4.4) has a measured $E(B-V)$ value of 0.12. Mk35, $\sim 10$\as south
from both R136 and Mk34 (Malumuth \& Heap 1994), has a value of 0.27.
{}From the conversion $N_H/E(B-V) = 2.4 \times 10^{22}
{\rm~cm^{-2}~mag^{-1}}$ (Fitzpatrick 1985),
$N_H \approx$ 2.9 and $6.5 \times 10^{21}
{\rm~cm^{-2}}$ for the two stars. Assuming that the stars are located in
the same regions as the corresponding X-ray sources, one
can then compare the estimated interstellar $N_H$ values with the column
densities from the spectral fits (Table 2). For example, the
MBD fits suggest considerable amounts of intrinsic absorption
($\gtrsim 2 \times 10^{21} {\rm~cm^{-2}}$) in both sources.
Such intrinsic absorption is expected for
wind-driven compact X-ray sources (e.g., Moffat \& Seggewiss 1979).

\subsection {Accretion Disk Model}

	WR+BH binaries provide a natural explanation of the observed X-ray
properties of the two sources. With their X-ray luminosities $\lesssim 10^{37}
{\rm~ergs~s^{-1}}$, the sources are likely driven by material
captured from massive winds of the WR stars (Conti 1978). The specific
angular momentum in the wind material can result in accretion
disks close to the BHs. The disks manifest themselves as ultrasoft components
in the X-ray spectra of the sources (e.g., Tanaka \& Lewin 1995). The
observed steep spectra in the
1-2~keV band do indicate the presence of optically-thick disks, and can
be approximately modeled with MBD (Table 3; \S~3.2).

	In comparison, an HMXB containing a strongly
magnetized neutron star ($B \sim 10^{12}$~G) does not
normally produce a steep soft X-ray spectrum at a
low luminosity state. Because of the strong magnetic field,
which is expected for
a young neutron star of an age comparable to or smaller than
that of R136, no X-ray-emitting accretion disk can be formed.
The X-ray emission from the star's surface should have a hard spectrum,
typically characterized by a power law of a photon index 1.0-2.0 up to
a high energy cutoff at 10-20~keV (e.g., White, Swank, \& Holt 1983).
This index is significantly smaller than those of Sources 5 and 7
(Table 2). Therefore, the sources do not appear to be neutron
star systems.

	Assuming MBD, one can estimate the mass
$M_x$ of an accreting compact object. If the disk inner radius
$R_{in}\sim 3 R_s$ (with $R_s = 2GM_x/c^2$ the Schwarzshild radius;
Ebisawa et al. 1991 and references therein),
where circular orbits become dynamically unstable, $M_x \gtrsim 2.5
{\rm~M_\odot/(cos}~i)^{0.5}$ for Source 5, and $= 4.3_{-3.2}^{+20}
{\rm~M_\odot/(cos}~i)^{0.5}$ for Source 7. These mass estimates are probably
underestimated due to the oversimplification of the MBD model.
The characteristic temperature ($T_{in}$) at the inner radius of a disk,
as given Table 2, is the color temperature rather than the effective
temperature, which could be considerably higher (by a factor $\sim 1.5$;
Ebisawa et al. 1991 and references therein). With this effect
in the consideration, the estimates of $R_{in}$, thereof $M_x$,
would be increased  by a factor of $\sim 2$ (Tanaka 1992). Although the
theoretical upper limit for a neutron star is $\sim 2.5 {\rm~M_\odot}$
(e.g., Shapiro \& Teukolsky 1983), radio observations have constrained
masses of neutron stars in binary pulsar systems to be within the
limits of 1 and 1.6${\rm~M}_\odot$ (e.g., Finn 1994). Thus, the
compact objects in Sources 5 and 7
are more  likely stellar mass BHs than neutron stars.

	Now, let us see whether or not the X-ray luminosities of
the sources can be explained by the BH accretion from stellar winds
of the WR stars. The bolometric luminosities ($L_{bol}$)
of Sources 5 and 7 are $\sim 4$ and $1 \times 10^{36} {\rm~ergs~s^{-1}}$,
estimated from the best-fit MBD parameters of each source (Table~2).
Further assuming $L_{bol} \approx 0.06 \dot{M} c^2$ (Sunyaev \& Titarchuk
1980), one can obtain the mass accretion rates ($\dot{M}$) as $\sim 10$
and $3 \times 10^{-10} {\rm~M_\odot~yr^{-1}}$. Theoretically, one can expect
an accretion rate as \begin{equation}
\sim (3 \times 10^{-10} {\rm~M_\odot~yr^{-1}}) ({V_{WR} \over 3 \times 10^3
{\rm~km~s^{-1}}})^{-4} ({\dot{M}_{WR} \over 10^{-5} {\rm~M_\odot~yr^{-1}}})
({R \over 15 {\rm~R_\odot}})^{-2} ({M_x \over 10 {\rm~M_\odot}})^2,
\end{equation} where $V_{WR}$ and $\dot{M}_{WR}$ are the wind velocity and
mass loss rate of a WR star, and $R$ is the distance between a WR star and
its BH component (Moffat et al. 1997). Within the uncertainties of the
parameters (Usov 1992 and references therein), the wind-driven accretion
scenario provides a reasonable explanation of the $\dot{M}$ values of the
two sources.

\subsection { Comparison with Known Black-Hole Candidates}

	Table 3 lists parameters of  the three previously known HMXB BHCs
(Cyg X-1, LMC X-3, and LMC X-1) as well as Sources 5 and 7.
The spectrum of an HMXB BHC can  be characterized in general by
a power law plus an ultrasoft component (e.g., MBD). Normally, the
power-law contribution dominates only at energies greater
than a few keV (outside the {\it ROSAT} band), and typically has a
photon index of $\sim 2.0 - 2.5$ (e.g., Schlegel et al.
1994; Tanaka \& Lewin 1995). Only Cyg X-1 often show a
`low' state in which the spectrum can be reasonably well fitted by a
single flatter power law of a photon index $\sim 1.5$ from
$\sim 2$ to 100~keV
(Liang \& Nolan 1984; Marshall et al. 1993; Tanaka \& Lewin 1995).
However,
an ultrasoft excess below $\sim 2$~keV is probably still required to
explain the discrepancy between
X-ray-absorbing gas column densities inferred from optical and X-ray
measurements (e.g., Balucinska \& Hasinger 1991).

	The evidence that the two 30 Doradus sources have a
power-law component expected for HMXB BHCs comes from a
 recent {\it ASCA} observation (Itoh et al. 1994), which is sensitive to
the 0.5 - 10~keV range. I have examined images constructed with the
observation in different energy bands. Counts in the 2 - 10~keV image
strongly concentrate to the central region where the two sources
are located (see also Itoh et al. 1994), whereas the image in
the 0.5 - 2~keV band shows a count distribution similar to that in Fig.~2.
Interestingly, the integrated spectrum of 30 Doradus can be well fitted with
two spectral components: an optically-thin thermal plasma and
a power law. These results suggest that the thermal plasma contribution,
dominating at energies $\lesssim 2$~keV, is overwhelmed by
diffuse hot gas, and that the  power-law component,
important at higher energies, can be explained, at least partially,
by the presence of the two sources. The photon index $2.42_{-0.19}^{+0.18}$,
given by Itoh et al. (1994), is consistent with the expected value for
HMXB BHCs. The luminosity of the power-law component is
$\sim 1.4 \times 10^{36} {\rm~ergs~s^{-1}}$ in the 0.5 - 10 keV range.
Even in the 0.5 - 2 keV range, the power-law contribution
of $\sim 8 \times 10^{35} {\rm~ergs~s^{-1}}$
may not be negligible, compared to the summed luminosity
of Sources 5 and 7 ($\sim 13 \times 10^{35} {\rm~ergs~s^{-1}}$), inferred
from the {\sl ROSAT} observations. However, the power-law contribution
could have been substantially overestimated. The fit
with a plasma of a single temperature
is probably an oversimplification; the use of a multi-temperature
plasma would reduce the index, thereof the contribution, of the power
law in the 0.5 - 2~keV band.

	The ultrasoft component of the sources,
revealed by the {\sl ROSAT} PSPC spectra, is another important signature
of HMXB BHCs (Tanaka 1994). From Table 3, one can find a general correlation
between $T_{in}$ and $L_x~(\sim L_{bol})$, which roughly agrees with
the relation predicted from  the MBD model (e.g., Ebisawa et al. 1991):
\begin{equation}
T_{in} \approx 0.6{\rm~keV} ({M_x/10{\rm~M_\odot}})^{-0.5}
(L_{bol}/10^{38} {\rm~ergs~s^{-1}})^{0.25},
\end{equation}
within the uncertainties of the relevant parameters. This indicates that the
ultrasoft spectra of the  sources can be explained by their relatively low
$L_x$ (or $\dot M$) values. The relatively low mass accretion rates  of
these two sources  are expected because they are most likely wind-driven
systems, whereas the known HMXB BHCs are powered by
Roche-lobe overflows (McClintock 1991 and references therein).
Moreover, the $R_{in}$ values of the sources (Table 3) are also consistent
with those of other HMXB BHCs. By contrast, low-mass X-ray binary BHCs
have $R_{in} \lesssim 10$~km and $T_{in}$ in the range
of 1.4 - 1.5~keV (Tanaka \& Lewin 1995), incompatible with the parameters
obtained for the sources.

	It, therefore, appears that the 30 Doradus sources represent an
extension of the known HMXB BHCs to low luminosities, and are
likely powered by accretion from stellar winds.

\subsection {Optical Properties}

The positional coincidence of Source 5 with R140 makes the WR+BH binary
explanation particularly attractive. R140 is a visual multiple star system.
The brightest three stars, which form the north-east elongation as
seen in Fig.~3, have been studied both
photometrically and spectroscopically by Moffat et al. (1987).
R140a1 (WC5) looks brighter than  normal, suggesting a possible
companion O star. But no variation in the radial velocity (RV) of the WR star
has been detected. R140b is likely a single normal WN6 star. Most
interesting is R140a2 (WN6), which is a close spectroscopic binary with a
period of 2.76 days. The nature of the binary partner is, however, unknown.
The optical light curve of the binary shows  phase-dependent
variations with an amplitude  $\gtrsim 0.04$ mag. But the optical colors
do not seem to vary. This could be explained by
electron scattering effects in the WR wind. From the observed intensity
changes and the lack of color variations,  Moffat et al. estimated that
the orbital inclination $i$  lies in the range of $30^\circ - 70^\circ$.
The orbit eccentricity $e$ is also  unknown. It is expected
to be close to zero if the binary of such a short period contains
two normal stars (Cherepashchuk 1991). However, if a supernova happened in
the binary, the orbit could be highly eccentric. But, this also depends on
whether or not there was considerable amounts of mass ejection during
the supernova explosion, and on how effective the tidal evolution may have
been in circularizing the orbit (Savonije \& Papaloizou 1984). An orbit of
large eccentricity should cause a noticeable phase-dependent
variation in $\dot{M}$ (\S~4.2), thereof, the source intensity.
The lack of such a variation in the observations may indicate
$ e \sim 0$. Furthermore, from the RV amplitudes of
\ion{He}{2} $\lambda4686$ and \ion{N}{4} $\lambda4058$ lines
of the WR star, Moffat et al. derived a mass function $f(m) = 0.10
{\rm~M}_\odot$.

	These estimates of the orbital parameters provide constraints
on the  mass  of the unknown binary partner of R140a2. From the definition
of the mass function, I obtain a solution
$ M_x = M_{WN}/[(M_{WN}/f(m))^{1/3} {\rm~sin}~i - 2/3]$.
If the mass of the WN star ($M_{WN}$) is within the known mass range  of
WN stars from 8 to 48${\rm~M}_\odot$ (Cherepashchuk 1991), the
solution gives $2.4 {\rm~M}_\odot < M_x < 15 {\rm~M}_\odot$,
consistent with the mass estimate from the MBD
fit to the PSPC spectrum of Source 5 (\S~4.2). Hence, it is reasonably
to assume that Source 5 is associated with R140a2 and
represents a BHC.

	Mk34 does not appear as a visual multiple star system in
an {\sl HST} WFC2 image. (R140 has not been observed by the {\sl HST}.)
Spectroscopic observations are
needed to see whether or not this WR star is also in a close binary system
similar to R140a2.

\subsection {Alternative Explanation: Colliding Wind Systems?}

	Could the two X-ray sources be explained as
 colliding wind systems? Or could the two WR stars be in binaries
containing only normal massive stars (e.g., WR+O)?
Theoretically, it may be possible to postulate a colliding wind
model accounting for an X-ray luminosity of up to $\sim 10^{35}
{\rm~ergs~s^{-1}}$ (e.g., Luo \& McCray 1990; Usov 1992). Observationally,
however, there is no evidence for such a luminous colliding wind system.
All WR stars, including binaries, in the Milky Way have their individual X-ray
luminosities $\lesssim 10^{34} {\rm~ergs~s^{-1}}$ (e.g., Pollock 1987).
No other WR star is known as an X-ray source in the LMC (Wang et al. 1991)
or in the SMC (Wang \& Wu 1992). About 40\% of LMC WR stars are in binaries
which could be colliding wind systems. A few of these binaries
have orbital periods comparable to or smaller than that of R140a2;
for instance, HD 36521 (WC4+O6V-III) has a period of 1.9 days
(Moffat et al. 1986). The non-detection of these binaries as X-ray sources
suggests that colliding wind systems in the LMC cannot be much brighter than
their Galactic counterparts. Furthermore, the binary partner of
R140a2 is unlikely a normal massive star. The upper bound of $M_x$ is
considerably smaller than the expected mass of an O star ($\gtrsim 20
{\rm~M}_\odot$). Also for the partner to be
an O star, the mass function requires $M_{WN}/M_x \gtrsim 4$, which falls
outside the range of 0.17 - 2.67 for all known WR+O binaries
(Cherepashchuk 1991). A binary containing two WR stars, or a WR star
accompanied by a supergiant, is very rare, since it arises
only from a wide pair with very close initial masses of components.
Therefore, colliding wind systems do not appear as a viable
explanation of the X-ray sources.

\section {Summary}

	{\sl ROSAT} observations have shown the presence of
two X-ray sources in the central region of 30 Doradus. These two sources
coincide in position with WR stars to $\lesssim 2^{\prime\prime}$.
One of the WR stars is in a spectroscopic binary system;
the unknown binary component with its mass in the range of 2.4 - 15~M$_\odot$
is likely a black hole, accreting from the stellar wind of the star.
Both sources show extremely steep spectra and possible intrinsic absorption
in the 0.5 - 2~keV range. Modeled with MBD, the spectra give $R_{in}$
as $\gtrsim 22$ for one source and  $= 38_{-18}^{+178}
{\rm~km/(cos}~i)^{0.5}$ for the other. These $R_{in}$ values
are comparable to those obtained for the known HMXB BHCs.
The ultrasoft characteristics of the spectra with $T_{in} \sim 0.2$~keV
is consistent with the relatively low luminosities
of the sources ($\sim 10^{36} {\rm~ergs~s^{-1}}$) compared
to the known BHCs. At energies $\gtrsim 2$~keV, covered by an {\sl ASCA}
observation, there is evidence for the presence of a power-law component with
a photon index of $\sim 2.4$, similar to the values of the HMXB BHCs.
Furthermore, the sources do not appear to be transients although aperiodic
intensity variability of a factor $\lesssim 2$
cannot be ruled out by the data. These results suggest that the two
X-ray sources are WR+BH binaries, most probably wind-driven systems.

	The confirmation of the two sources as WR+BH binaries
at a known distance of the LMC will have
important implications for our understanding of the evolution and
mass distribution of massive stars (e.g., the calculation
of the initial mass function of 30 Doradus stars;  Parker \& Garmany
1994; Cherepashchuk 1991 and references therein) as well as for
studying accretion processes of compact objects.

	There might be similar sources in the Milky Way, especially in recent
star formation regions. But, the ultrasoft component of
such sources would be difficult to detect because of
severe line-of-sight absorption of soft X rays. In the LMC, the interstellar
absorption is relatively smaller, partly because of the lower metallicity of
gas in the Cloud. Therefore, it is easier to detect sources with  ultrasoft
spectra in the LMC than in the Milky Way.

\acknowledgements

	I thank J. Parker for providing the CCD image and
astrometry information, A. F. J. Moffat for commenting on
 optical observations of Wolf-Rayet stars, and the
referee for comments toward the improvement of this paper. I am grateful
to G. S. Miller and R. Taam for discussing the black-hole scenario and
for commenting on an early version of the paper.
This work  was supported by the Lindheimer Fellowship and NASA grant
NAG5-2717.

\begin{table}[htb]
\scriptsize
\begin{tabular}{lccccrll}
%\tablewidth{8in}
\multicolumn{8}{c}{\bf Table 1}\\
\multicolumn{8}{c}{\bf {\sl ROSAT} HRI Detections of Discrete X-ray Sources in
the 30 Doradus
field\tablenotemark{a}}\\ [0.1in]
\hline \hline
Source & RA & Decl. &  RA & Decl. & \multicolumn{1}{c}{HRI Rate} & & \\
No. &\omit\span (X-ray) &\omit \span (optical) & \multicolumn{1}{c}{(counts
ks$^{-1}$)} & \ \ ID & Comments and
References\tablenotemark{b}\\
%\documentstyle{apjpt}
%\begin{table}[htb]
%\tablecaption{{\sl ROSAT} HRI sources in the 30 Dor field\tablenotemark{a}}
%\begin{planotabule}{lccccl} \hline \hline
%\tablehead{Source\# & R.A. & Decl. &HRI Rate\ \ \ &}
%\tablehead{&(J2000) & (J2000) & (counts ks$^{-1}$) &  ID & Comments
%%\tablenotemark{b}
%tablewidth{0pt}
\hline
  1  &   5$^h$37$^m$47\fs6 & -69\deg10\arcmin20$^{\prime\prime}$& & &     52.5
$\pm$    1.4 &SNR N157B &p(?)+d, [1]\\
 2  &   5  38  10.2 & -68  57  01 &  & &    2.5 $\pm$    0.5 & &p\\

 3 &   5  38  16.7 & -69  23  32 & 5 38 16.3 & -69 23 34  &42.6 $\pm$ 1.6 & cal
69 &p, Galactic star, dK7e, [2]\\

 4 &   5  38  34.7 & -68  53  07 &  5 38 34.6 & -68 53 07 &   69.5 $\pm$    1.8
& cal 71 &p, Galactic star, G2V, [2]\\

 5  &   5  38  41.8 & -69  05  14 & 5 38 41.7 & -69 05 13.5 &     5.1 $\pm$
0.5 & R140 &p, WR, [3] [4]\\
 6\tablenotemark{c}  &   5  38  44.0 & -68  52  42 & & &    10.9 $\pm$    0.9 &
& p\\
 7  &   5  38  44.4 & -69  06  07 & 5 38 44.3 & -69 06 05.3 &      8.1 $\pm$
0.6 & Mk34 & p+d, WN4.5,  [3] [4]\\

 8 &   5  40  11.1 & -69  19  55 & 5 40 11.0 & -69 19 55.2 &  239.9 $\pm$
3.2 & SNR N158 &p+d, 50ms pulsar, [5] [6]\\ \hline
\end{tabular}
\newpage
\tablenotetext{a}{The equinox of the positions is J2000.
Peaks that represent intensity enhancements of diffuse X-ray features
are not included. The optical positions should be accurate to within
$\sim 1{\prime\prime}$.
The uncertainties in the X-ray positions are
primarily systematic, ranging from $\lesssim 3^{\prime\prime}$
on axis to $\sim 10^{\prime\prime}$ at $\sim 15^{\prime}$. }

\tablenotetext{b}{A letter ``p'' (or ``d'') indicates the presence of a
point-like (or diffuse) component of a source. References are for optical
positions and
IDs: [1] Dickel et al. 1993; [2] Cowley et al. (1984); [3]
Moffat et al. (1987); [4] Parker (1993); [5] Shearer et al. (1994) ; [6]
Seward \& Harnden Jr. (1994).}

\tablenotetext{c}{The count rate is uncertain because of confusion with
Source  4.}
\end{table}
\vfil
\eject

\begin{table}[htb]
\begin{tabular}{lrlrlc}
\multicolumn{6}{c}{\bf Table 2}\\
%\caption{Source Spectral Parameters\tablenotemark{a}} \\ [0.1in]
\multicolumn{6}{c}{\bf Source Spectral Parameters\tablenotemark{a}} \\ [0.1in]
\hline \hline
\multicolumn{1}{c}{Model} & \multicolumn{2}{c}{$T$ (keV) or $\alpha$} &
\multicolumn{2}{c}{$N_H$ ($10^{22} {\rm~cm^{-2}}$)} & $\chi^2$\\ \hline
& \multicolumn{5}{c}{ Source 5}\\
  BB      &   0.11  &(0.05, 0.18)& 1.3  &(0.4, 3.6) & 12\\
  MBD & 0.13 &(0.10, 0.21) & 1.2 &(0.5, 2.0) &12\\
  PL       &   10    &($\geq 6.5$) & 2.0  &($\geq 1.8$) & 12\\ [0.07in]
& \multicolumn{5}{c}{Source 7}\\
  BB     &   0.21 & (0.15, 0.32)& 1.2  &(0.54, 1.9) & 35\\
  MBD & 0.24 &(0.17, 0.35) & 1.4 &(0.95, 2.1) &35\\
  PL      &   5.6  &  (3.2, 9.0) & 1.8 & (1.0, 3.1) & 34\\ \hline
\end{tabular}
\tablenotetext{a}{\hsize=2truein  Parameters were included for
three spectral models:  black-body (BB), multicolor blackbody
 disk (MBD; Pringle 1981; Makishima et al. 1986), and power law (PL).
$\alpha$ is the photon index of a power law.
$N_H$ is the equivalent hydrogen column density of X-ray-absorbing gas with
assumed metal abundances as 40\% of the solar values
 (Meyer et al. 1994; de Boer et al. 1985; Morrison \& McCammon 1983).
Parameter limits  are all at the 90\% confidence level.
The number of degrees of freedom in the spectral fits is 17 (Source 5) and
18 (Source 7).}
%\end{planotable}
\end{table}
\begin{table}[htb]
\begin{tabular}{lccccc}
\multicolumn{6}{c}{\bf Table 3}\\
\multicolumn{6}{c}{\bf Comparisons of Black-Hole Candidates\tablenotemark{a}}
\\ [0.1in]
\hline \hline
\multicolumn{1}{c}{Source} & Source 5 &Source 7 & Cyg X-1 & LMC X-3 & LMC X-1\\
\hline
Orbital Period (d) & 2.8 &--& 5.6 & 1.7 & 4.2 \\
f(m) (${\rm M_\odot}$) & 0.10 &--& 0.25 & 2.3 & 0.14(?)\\
Companion & WN6 & WN4.5& O9.7Iab & B3V & O7-9III(?)\\
Transient & no &no& no & no & no\\
$L_x ({\rm~ergs~s^{-1}})$  & $\sim 10^{36}$ & $\sim
10^{36}$ & $\sim 2 \times 10^{37}$ & $\sim 4 \times 10^{38}$ & $\sim 2 \times
10^{38}$\\
$T_{in}$ (keV) & 0.10-0.21 & 0.17-0.35\tablenotemark{b} & $\sim$ 0.38 & $\sim$
1.1 & $\sim$ 0.83 \\
${\rm R_{in}(cos}~i)^{0.5} {\rm~(km)}$ & $\gtrsim 22$ &10-216& 18-30 (D/2.5
kpc) & $\sim 24$ & $\sim 40$ \\ \hline
\end{tabular}
\tablenotetext{a}{\hsize=2truein  Parameters for Cyg X-1, LMC X-1, and
LMC X-3 are from Ebisawa et al. (1991), Schlegel et al. (1994),
Tanaka (1994), and Tanaka \& Lewin (1995).
The candidacy of LMC X-1 is somewhat
questionable (Tanaka \& Lewin 1995 and references therein). X-ray
luminosities are estimated roughly in 1.2 - 10~keV range.
Optical parameters for Sources 5 and 7 are from Moffat et al. (1986). The
values of ${\rm R_{in} (cos}~i)^{0.5}$, where $R_{in}$
and  $i$ are the inner radius and inclination angle
of an accretion disk, are obtained from spectral fits with
the multicolor blackbody disk model (Makishima et al. 1986; Ebisawa et al.
1991).}
\tablenotetext{a}{Probably overestimated
because of the spectral contamination from R136 (\S~3.2).}
%\end{planotable}
\end{table}

\clearpage

\centerline{\bf Figure Captions}

\begin{description}
\item [Fig.~1:]  30 Doradus region in X ray and optical. Contours of
X-ray intensity, smoothed with
a Gaussian of FWHM $= 18^{\prime\prime}$, are at 2.5, 4, 7, 12, 20, 40, 80,
and 160$\sigma$ (1$\sigma = 1.0 \times
10^{-3} {\rm~counts~s^{-1}~arcmin^{-2}}$) above a local background of
$5.6 \times 10^{-3} {\rm~counts~s^{-1}~arcmin^{-2}}$. The gray-scaled image
is a representation
of a digitized ESO survey blue plate taken on December-3, 1979.

\item [Fig.~2:] X-ray image of 30 Doradus. The image is
adaptively smoothed with a Gaussian of adjustable size to achieve
a constant local count-to-noise ratio of $\sim$ 8 over the image.
The contours are at 3, 4, 6, 9, 13, 18, 24, 31, 40, 50, 100, 200, 400,
and 800$\sigma$ (1$\sigma = 5.0 \times
10^{-4} {\rm~counts~s^{-1}~arcmin^{-2}}$) above a local background of
$5.3 \times 10^{-3} {\rm~counts~s^{-1}~arcmin^{-2}}$.

\item [Fig.~3:] Central region of 30 Doradus in X ray and optical. The
X-ray intensity is a coadd of the two RHRI images and is smoothed with
a Gaussian of FWHM$=4^{\prime\prime}$. The lowest
contour is 3$\sigma$ (1$\sigma = 1.93 \times
10^{-3} {\rm~counts~s^{-1}~arcmin^{-2}}$) above a background level of
6 $\times 10^{-3} {\rm~counts~s^{-1}~arcmin^{-2}}$; the
intensity increment of
each subsequent contour is 1.5 times the previous one.
The optical CCD image was gray-scaled logarithmically. The
astrometry of the plate, corrected with star positions in
Parker (1993), should be better than $\sim 1^{\prime\prime}$.

\item [Fig.~4:] RHRI radial intensity distributions of Sources 5
({\sl triangles}) and  7 ({\sl squares}),
compared to the expected PSF of the RHRI system (solid curve).

\item [Fig.~5:] {\sl ROSAT} PSPC spectra of Sources 5 (a) and 7(b),
together with the best fits of the multicolor blackbody
disk model (Table 2).
\end{description}
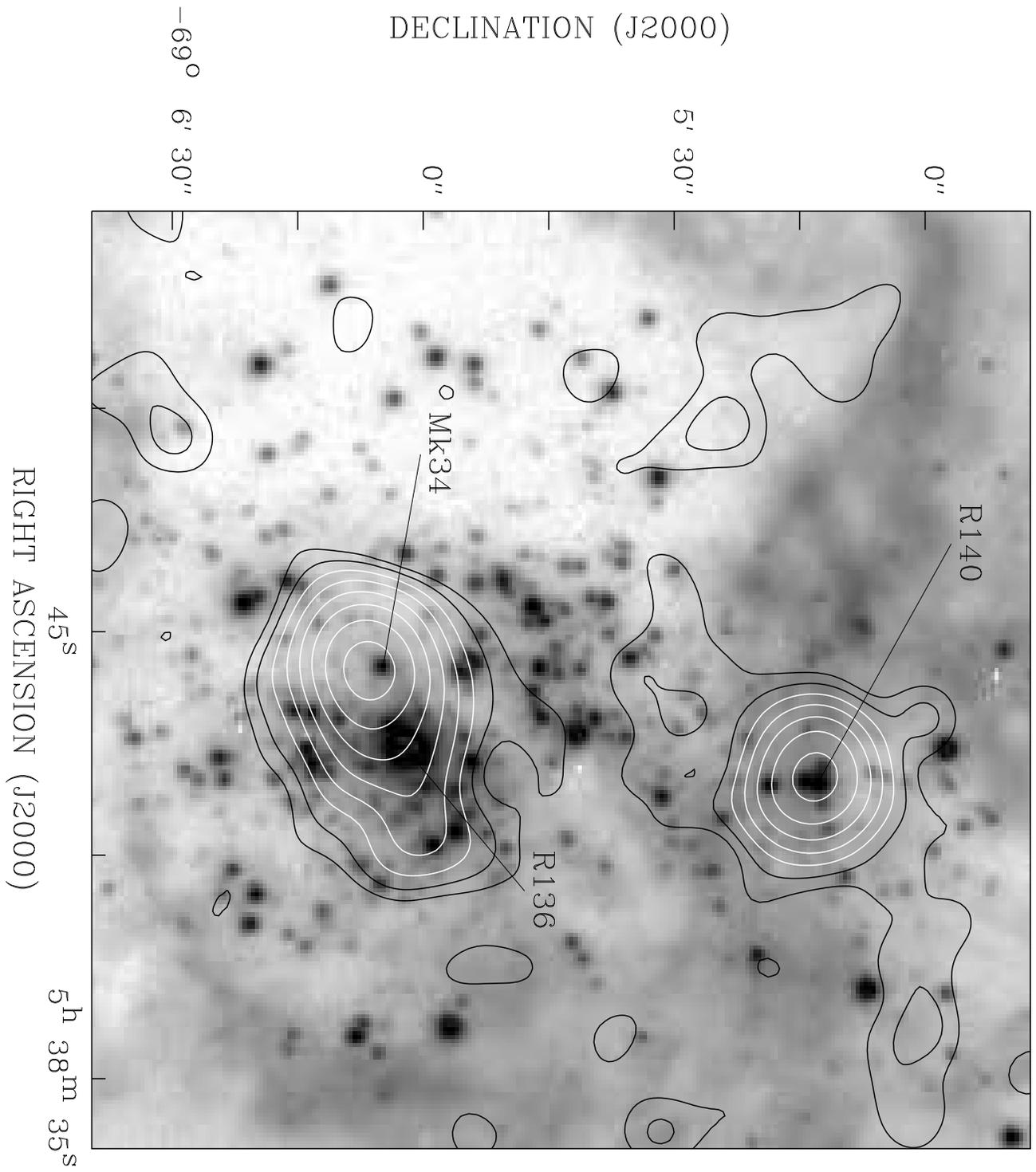
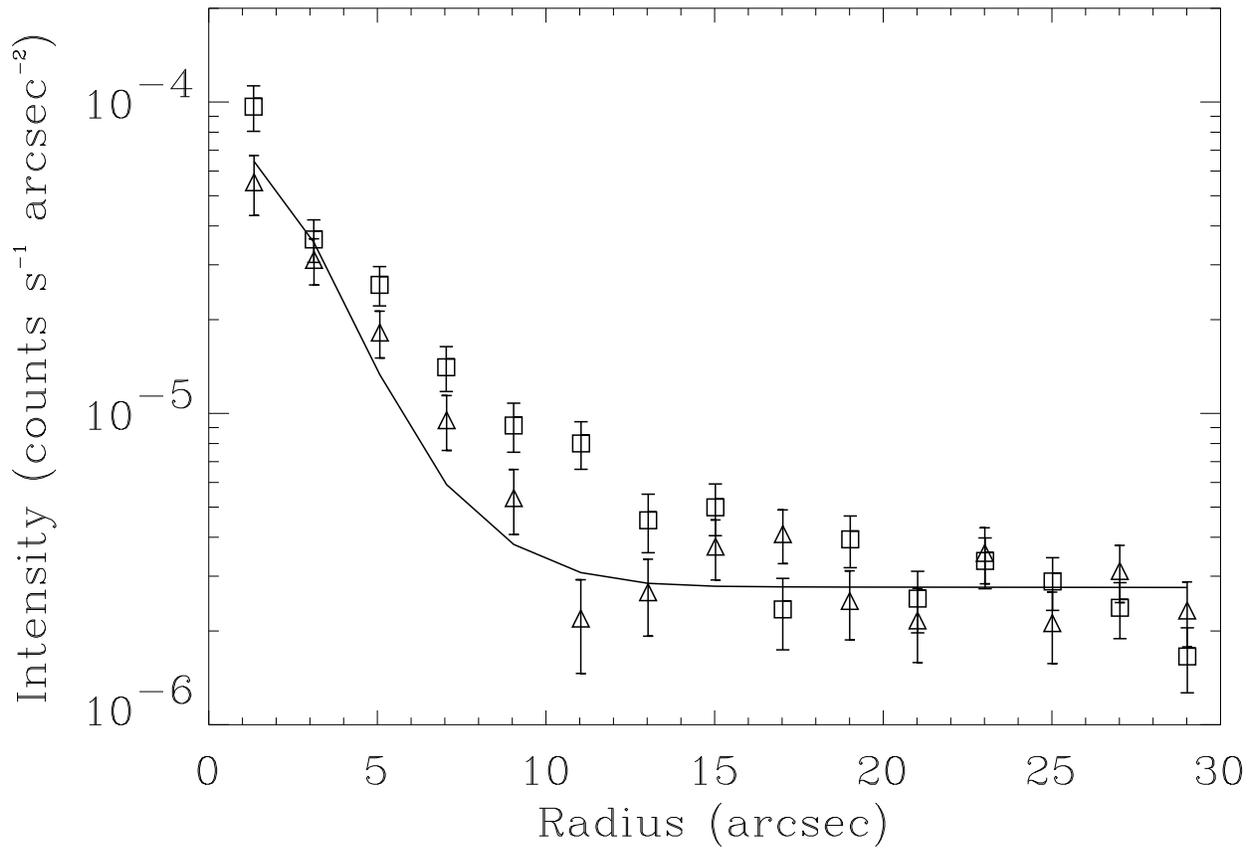
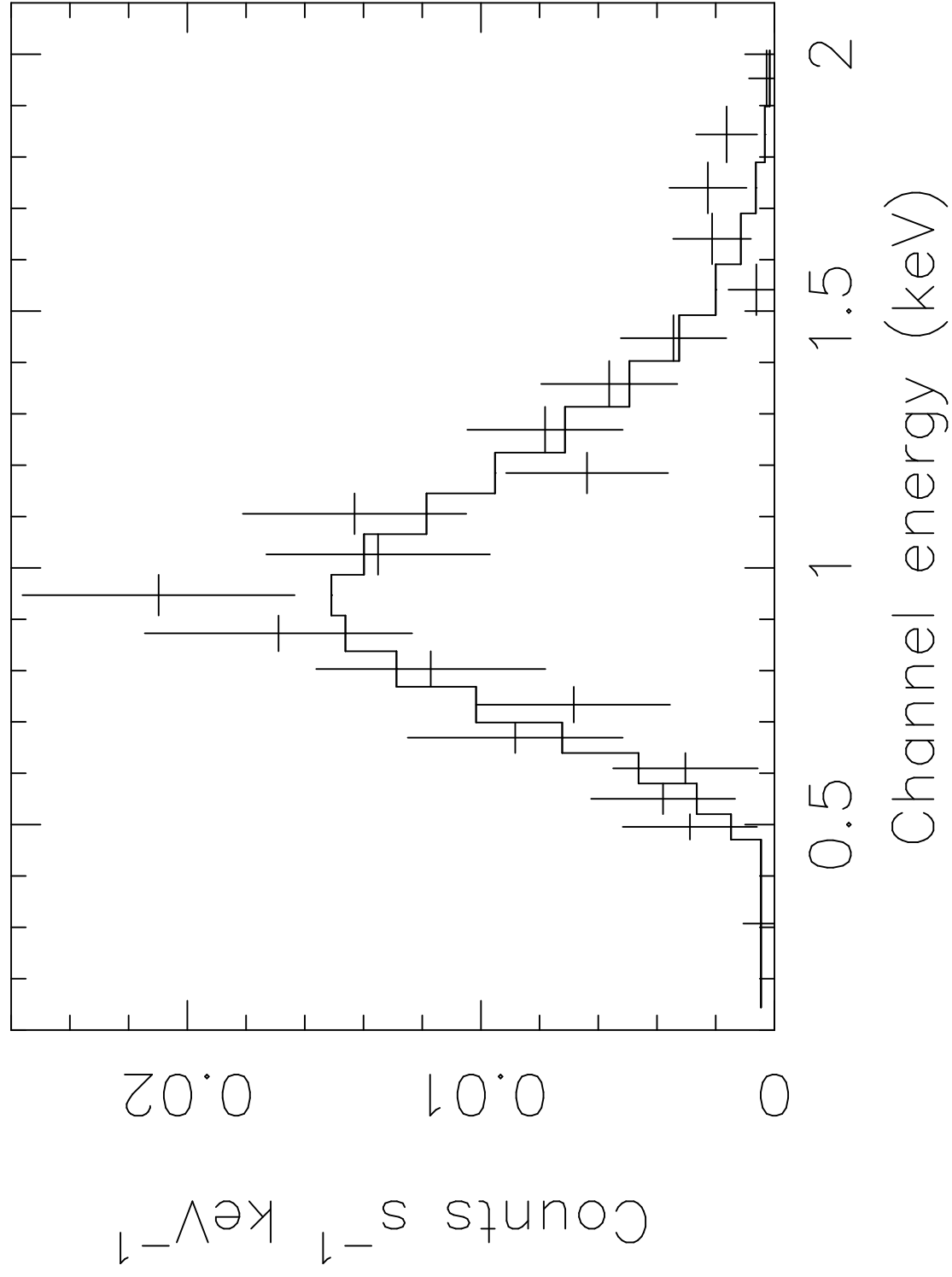
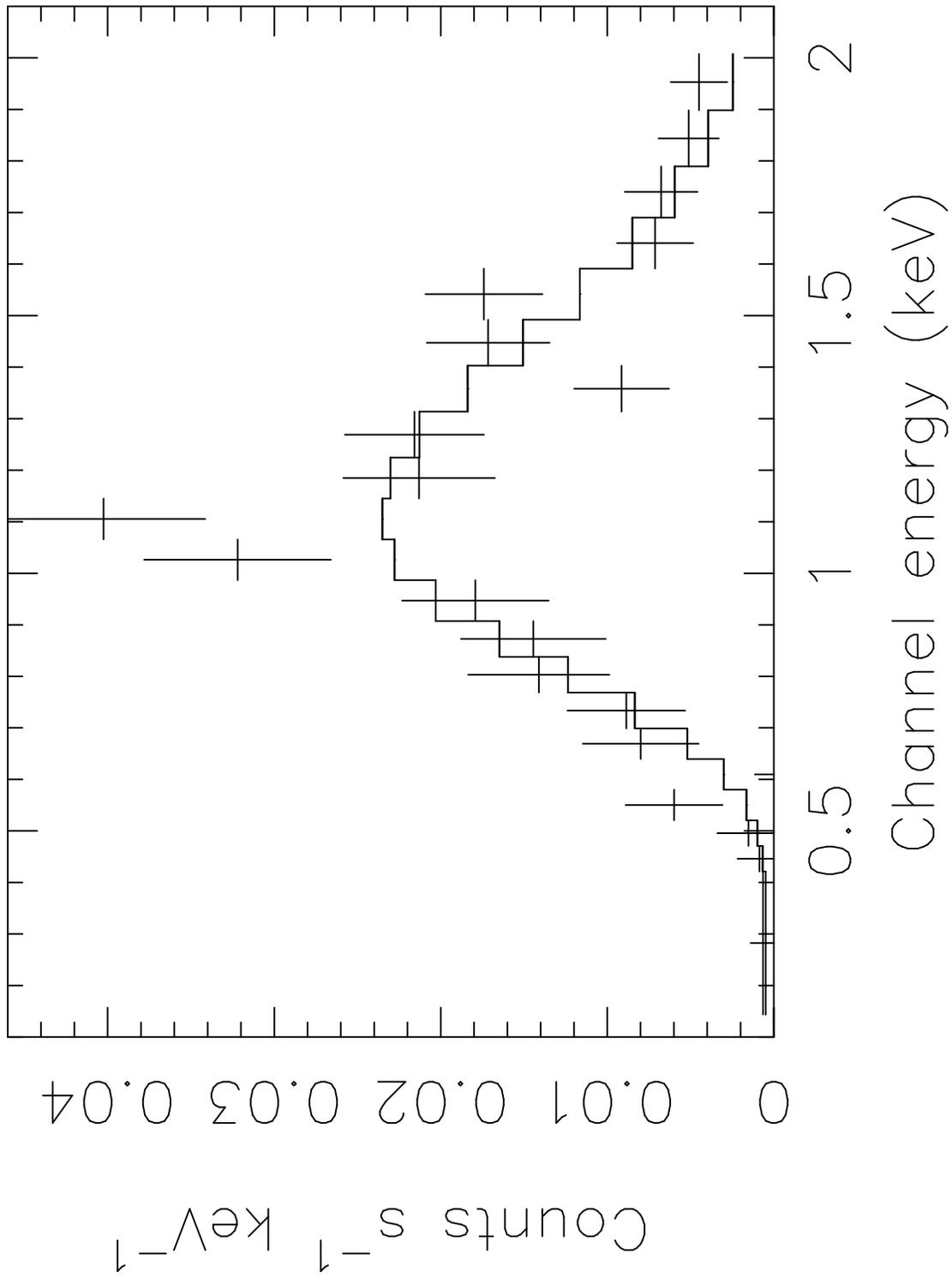
\end{document}